\documentclass[aps, twocolumn, superscriptaddress]{revtex4} 

\usepackage[switch]{lineno}
\usepackage{amsmath}
\usepackage{empheq}
\usepackage[parfill]{parskip}
\usepackage[active]{srcltx}
\usepackage{color}
\usepackage{array}
\newcolumntype{P}[1]{>{\centering\arraybackslash}p{#1}}
\usepackage{amsfonts}
\usepackage{dsfont}
\usepackage{graphicx}
\usepackage{subfig}
\begin{document}

\setlength{\parindent}{0.5cm}

\title{A review of swarmalators and their potential in bio-inspired computing}

\author{Kevin O'Keeffe}
\affiliation{Senseable City Lab, Massachusetts Institute of Technology, Cambridge, MA 02139} 
\author{Christian Bettstetter}
\affiliation{Institute of Networked and Embedded Systems, University of Klagenfurt, 9020, Austria}

\begin{abstract}
From fireflies to heart cells, many systems in Nature show the remarkable ability to spontaneously fall into synchrony. By imitating Nature's success at self-synchronizing, scientists have designed cost-effective methods to achieve synchrony in the lab, with applications ranging from wireless sensor networks to radio transmission. A similar story has occurred in the study of swarms, where inspiration from the behavior flocks of birds and schools of fish has led to 'low-footprint' algorithms for multi-robot systems. Here, we continue this 'bio-inspired' tradition, by speculating on the technological benefit of fusing swarming with synchronization. The subject of recent theoretical work, minimal models of so-called 'swarmalator' systems exhibit rich spatiotemporal patterns, hinting at utility in 'bottom-up' robotic swarms. We review the theoretical work on swarmalators, identify possible realizations in Nature, and discuss their potential applications in technology.
 \end{abstract}
 
 

\maketitle


\section{Introduction}
In 1967 Winfree proposed a model for coupled oscillators that spontaneously synchronize \cite{winfree2001geometry}. Beyond a critical coupling strength, the oscillators overcome the disordering effect of the dissimilarities in their natural frequencies and spontaneously lock their cycles. Kuramoto later simplified Winfree's model and solved it exactly \cite{kuramoto1975international}. Since then, the study of sync has matured into a vibrant field \cite{strogatz2004sync,pikovsky2003synchronization,acebron2005kuramoto}. On the theoretical side, theorists have twisted Kuramoto's model in various ways resulting in rich phenomena like glassy behavior \cite{daido1992quasientrainment, bonilla1993glassy,iatsenko2014glassy, ottino2018volcano} and chimeras states \cite{kuramoto2002coexistence,abrams2004chimera,abrams2008solvable}. On the applied side, coupled oscillators have found use in neurobiology \cite{montbrio2015macroscopic, pazo2014low, o2016dynamics,luke2013complete}, cardiac dynamics \cite{peskin1975mathematical, grudzinski2004modeling,gois2009analysis}, and the bunching of school buses \cite{saw2018bus}. An interesting application of sync is the design of `bio-inspired' algorithms. Here, by aping the simplicity of coupled oscillator models, researchers have devised resource-efficient algorithms to procure synchrony in the lab, useful for networked computing and robotics~\cite{werner-allen05,hong05,babaoglu2007firefly,tyrrell10}.

A story with many parallels to the sync story has evolved in the study of swarms. In 1995 Vicsek proposed a simple model of swarming agents \cite{vicsek1995novel}, which -- like the sync transition of coupled oscillators -- showed a transition from disorder to order: beyond a critical coupling strength, the agents switched from a gas-like, incoherent state to one in which the agents moved as a coherent flock. The novelty of the out-of-equilibrium nature of the flocking transition -- the system is out of equilibrium since agents constantly consume energy to propel themselves -- piqued the minds of physicists and other theorists, which in turn helped give rise to the field of active matter \cite{ramaswamy2010mechanics,marchetti2013hydrodynamics,sanchez2012spontaneous}, a field perhaps more vibrant than the field of synchronization. In a final mirroring of the sync story, the bio-inspired community has also mimicked the minimalism of Vicsek's and other models of swarming  to design novel algorithms for optimization \cite{dorigo2010ant,meng2014new,meng2016new,mirjalili2017salp,yang2013swarm,shi2001particle} and robotic swarms \cite{hamann2018swarm,schmickl2011beeclust,savkin2004coordinated,merkle2007swarm,yang2013swarm}.

These stories demonstrate swarming and synchronization are intimately related. In a sense, the two effects are `spatiotemporal opposites': in synchronization the units self-organize in time but not in space; in swarming the units self-organize in space but not in time. Given this conceptual twin\-ship, it is natural to wonder about the possibility of units which can self-organize in both space and time -- that is, to wonder how swarming and synchronization might interact. And more pertinent to this work, to wonder how a mix of swarming and synchronization could be useful in technology.

Several researchers have started to address these questions by analyzing systems that both sync and swarm. Von Brecht and Unimsky have generalized swarming particles by endowing them with an internal polarization vector \cite{von2016anisotropic}, equivalent to an oscillator's phase. Others have attacked the problem from the other direction, by considering synchronizing oscillators able to move around in space~\cite{uriu2013dynamics,sevilla2014synchronization,stilwell2006sufficient,frasca2008synchronization}. In these studies, however, the oscillators' movements affect their phase dynamics, but not the other way around; thus, the interaction between swarming and synchronization is only one-way. Two-way interaction between swarming and synchronization has also been considered: The pioneering work is the Iwasa-Tanaka model of chemotactic oscillators \cite{tanaka2007general,iwasa2010hierarchical}, i.e., oscillators which interact through a background diffusing chemical. More recent works were carried out by Starnini et al \cite{starnini2016emergence}, Belovs et al \cite{belovs2017synchronized}, and O'Keeffe et al \cite{o2017oscillators} who proposed `bottom-up' toy models without reference to a background medium. O'Keeffe et al called the elements of their systems `swarmalators', to capture their twin identities as swarming oscillators, and to distinguish them from the mobile oscillators mentioned above for which the coupling between swarming and synchronization is~unidirectional. 

\begin{figure*}[t]
 \includegraphics[width = 2 \columnwidth]{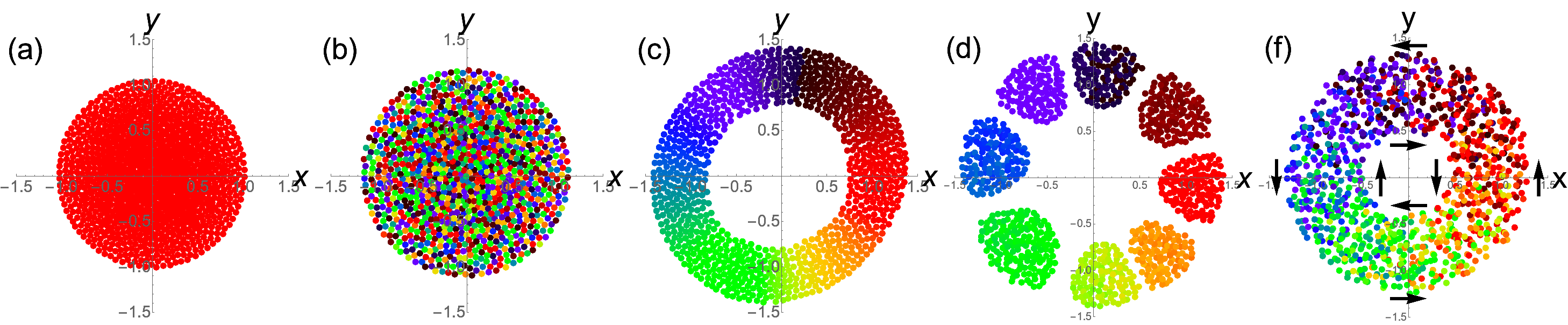}
 \caption{\textbf{Swarmalator states}. Scatter plots in the $(x,y)$ plane, where the swarmalators are colored according to their phase.  (a) Static sync for $(J,K) = (0.1, 1)$.   (b) Static async $(J,K) = (0.1, -1)$. (c) Static phase wave  $(J, K) = (1, 0)$.  (d) Splintered phase wave $ (J,K) = (1, -0.1) $. (e) Active phase wave $(J,K) = (1,-0.75)$.}
  \label{stationary_states_2d}
  \end{figure*}

This paper reviews research on swarmalators and other systems that mix swarming and synchronization. Our main motivation is to identify if the interplay of sync and swarming can be useful for bio-inspired computing and related engineering domains. We outline the theoretical work on swarmalators, experimental realizations, and finally conjecture on their technological utility.


\section{Models combining swarming and synchronization}
To combine synchronization with swarming, we first define what we mean by swarming. While to our knowledge there is no universally agreed on definition, swarming systems typically have at least one of two key features: ($i$) aggregation, arising from a balance between the units' mutual attraction and repulsion and ($ii$) alignment, referring to the units' tendency to align their orientation in space and move in a flock. Synchronization can thus be combined with either aggregation alone, alignment alone, or with both aggregation and alignment. In what follows, we present classes of model based on these three ways to combine sync and swarming.

\begin{figure} [h!]
 \includegraphics[width= 0.99 \linewidth]{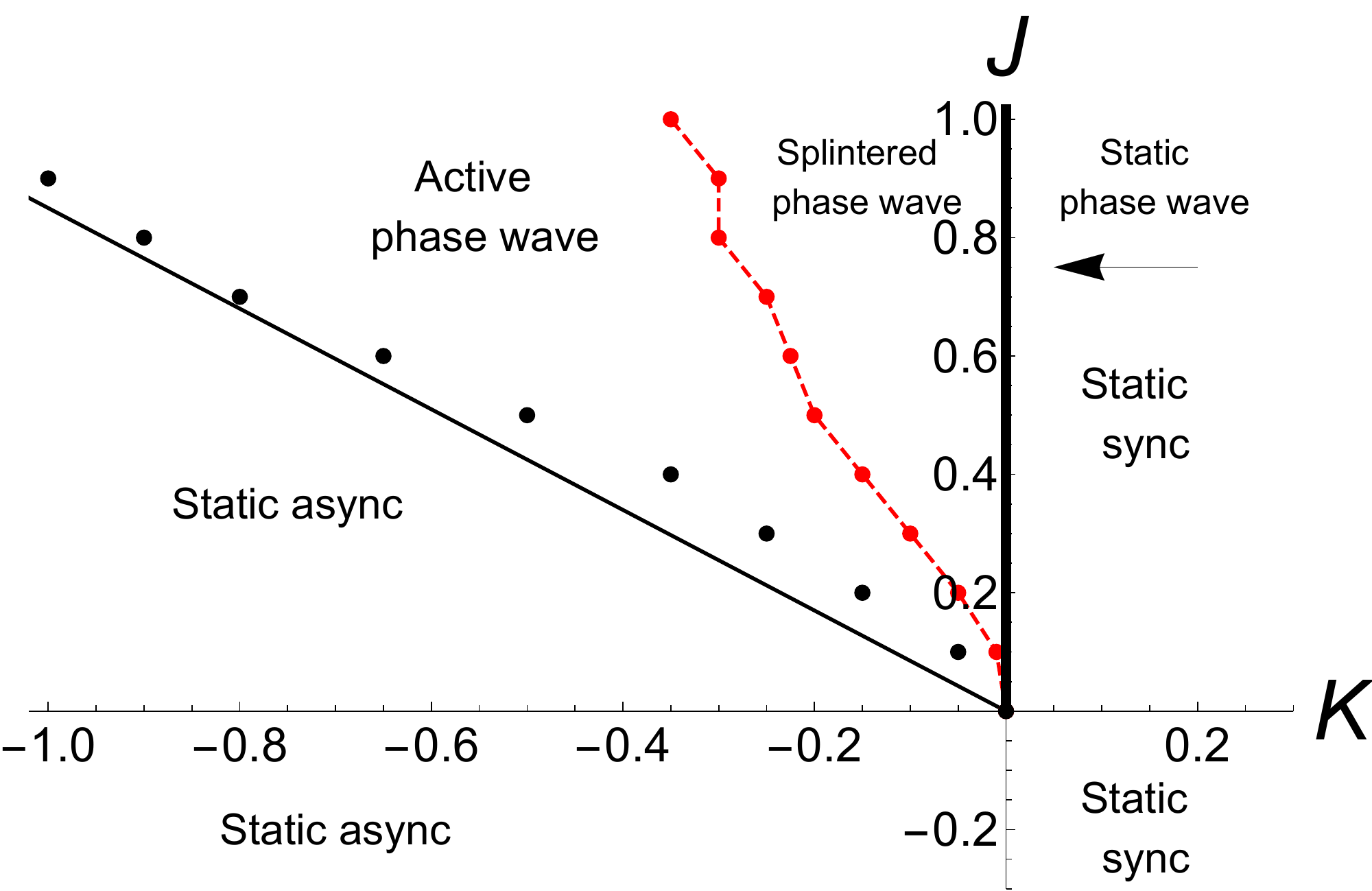}
  \caption{\textbf{Phase diagram}. Locations of states of the model defined by equations~\eqref{x_eom_model} and \eqref{theta_eom_model} in the $(J,K)$ plane. The straight line separating the static async and active phase wave states is a semi-analytic approximation given by $J \approx 1.2 K$ -- see Eq.~(18) in \cite{o2017oscillators}. Black and red dots show simulation data. The red dashed line simply connects the red dots and was included to make the boundary visually clearer. Note, this figure has been reproduced from \cite{o2017oscillators}.} 
 \label{phase_diagram_model1}
\end{figure}

\subsection{Aggregation and Synchronization}
The paradigmatic model of biological aggregation has the form
\begin{align}
&\dot{\mathbf{x}}_i = \frac{1}{N} \sum_{j \neq i}^N \mathbf{I}_{\rm att}( \mathbf{x_j} - \mathbf{x_i}) - \mathbf{I}_{\rm rep}( \mathbf{x_j} - \mathbf{x_i} ), \label{agg} 
\end{align}
\noindent
where $\mathbf{x}_i \in \mathbb{R}^d$ (usually with $d \leq 3$) is the $i$-th particle's position, $\mathbf{I}_{\rm att}$ captures the attraction between particles, and $\mathbf{I}_{\rm rep}$ captures the repulsion between them. The competition between  $\mathbf{I}_{\rm att}$ and $\mathbf{I}_{\rm rep}$ gives rise to congregations of particles with sharp boundaries, in accordance with many biological systems (see \cite{topaz2004swarming,mogilner1999non} for a~review).

The paradigmatic model in synchronization is the Kuramoto model \cite{kuramoto91}:
\begin{align}
\dot{\theta_i} = \omega_i + \frac{K}{N} \sum_{j \neq i}^N \sin(\theta_j - \theta_i). \label{kuramoto}
\end{align}
\noindent
Here $\theta_i \in S^1$ and $\omega_i$ are the phase and natural frequency of the $i$-th oscillator. The sine term captures the oscillators' coupling, where $K > 0$ means oscillators tend to synchronize, and $K < 0$ means oscillators tend to desynchronize. As mentioned, beyond a critical coupling strength $K_c$, a fraction of oscillators overcome the disordering effects imposed by their distributed natural frequencies $\omega_i$ and spontaneously synchronize.

To study the co-action of synchronization and aggregation, a natural strategy would be to stitch the aggregation model \eqref{agg} and the Kuramoto model \eqref{kuramoto} together. This was the approach taken by O'Keeffe et al \cite{o2017oscillators} who proposed the following swarmalator model:
\begin{align}
&\dot{\mathbf{x}}_i = \frac{1}{N} \sum_{ j \neq i}^N \Bigg[ \frac{\mathbf{x}_j - \mathbf{x}_i}{|\mathbf{x}_j - \mathbf{x}_i|} \Big( 1 + J \cos(\theta_j - \theta_i)  \Big) -   \frac{\mathbf{x}_j - \mathbf{x}_i}{ | \mathbf{x}_j - \mathbf{x}_i|^2}\Bigg]\label{x_eom_model} \\ 
& \dot{\theta_i} = \omega_i + \frac{K}{N} \sum_{j \neq i}^N \frac{ \sin(\theta_j - \theta_i)}{ |\mathbf{x}_j - \mathbf{x}_i| } . \label{theta_eom_model}
\end{align}
\noindent
Equation~\eqref{x_eom_model} models phase-dependent aggregation and Equation~\eqref{theta_eom_model} models position-dependent synchronization. The interaction between the space and phase dynamics is captured by the term  $1 + J \cos(\theta_j - \theta_i) $. If $J > 0$, ``like attracts like'': swarmalators are preferentially attracted to other swarmalators with the same phase, while $J < 0$ indicates the opposite. For simplicity the authors considered identical swarmalators $\omega_i = \omega$, and by a change of reference they set $\omega = 0$. 

The swarmalator model exhibits five long-term collective states. Figure~\ref{stationary_states_2d} showcases these states as scatter plots in the $(x,y)$ plane, where swarmalators are represented by dots and the color of each dot represents the swarmalator's phase $\theta$ (color, recall, can be mapped to $S^1$ and so can be used to represent swarmalators' phases). The parameter dependence of these states are encapsulated in the phase diagram shown in Figure~\ref{stationary_states_2d}. The first three states, named the \textit{static sync}, \textit{static async}, and \textit{static phase wave}, are -- as their names suggest -- static in the sense that the individual swarmalators are stationary in both position and phase. This stationarity allows the density of these swarmalators $\rho(\mathbf{x}, \theta)$ in these states to be constructed explicitly (the density $\rho(\mathbf{x}, \theta)$ is interpretted in the Eulerian sense, so that $\rho(\mathbf{x} + \mathbf{dx}, \theta + d \theta)$ gives the fraction of swarmalators with positions between $\mathbf{x}$ and $\mathbf{x} + d \mathbf{x}$ and phases between $\theta$ and $\theta + d \theta$). In the remaining \textit{splintered phase wave} and \textit{active phase waves} states swarmalators are no longer stationary. In the splintered phase wave state swarmalators execute small oscillation in both space and phase within each cluster. In the active phase wave the swarmalators split into counter-rotating groups -- in both space and phase --  so that $ \langle \dot{\theta_i} \rangle = \langle \dot{\phi_i} \rangle = 0$, where $\phi_i = \arctan(y_i / x_i)$ is the spatial angle of the $i$-th oscillator and angle brackets denote population averages. The conservation of these two quantities follows from the governing equations; the pairwise terms are odd and thus cancel under summation. Three-dimensional analogues of the five collective states were also reported.

The stability properties of the static async state are unusual. Via a linear stability analysis in density space, an integral equation for the eigenvalues $\lambda$ was derived. Yet numerical solutions to this integral equation, in the parameter regime where the static async state should be stable, produced a leading eigenvalue so small in magnitude that its sign could not be determined reliably. Thus, the stability of the state could not be analytically confirmed. There is however a parameter value at which the magnitude of $\lambda$ increases sharply, which was used to derive a pseudo critical parameter value $K_c \approx - 1.2 J$ marking the apparent (i.e.~as observed in simulations) destabilization of the state. Nevertheless, the true stability of the static async state remains a puzzle.

\begin{figure}[h]
 \includegraphics[width= 0.60 \columnwidth]{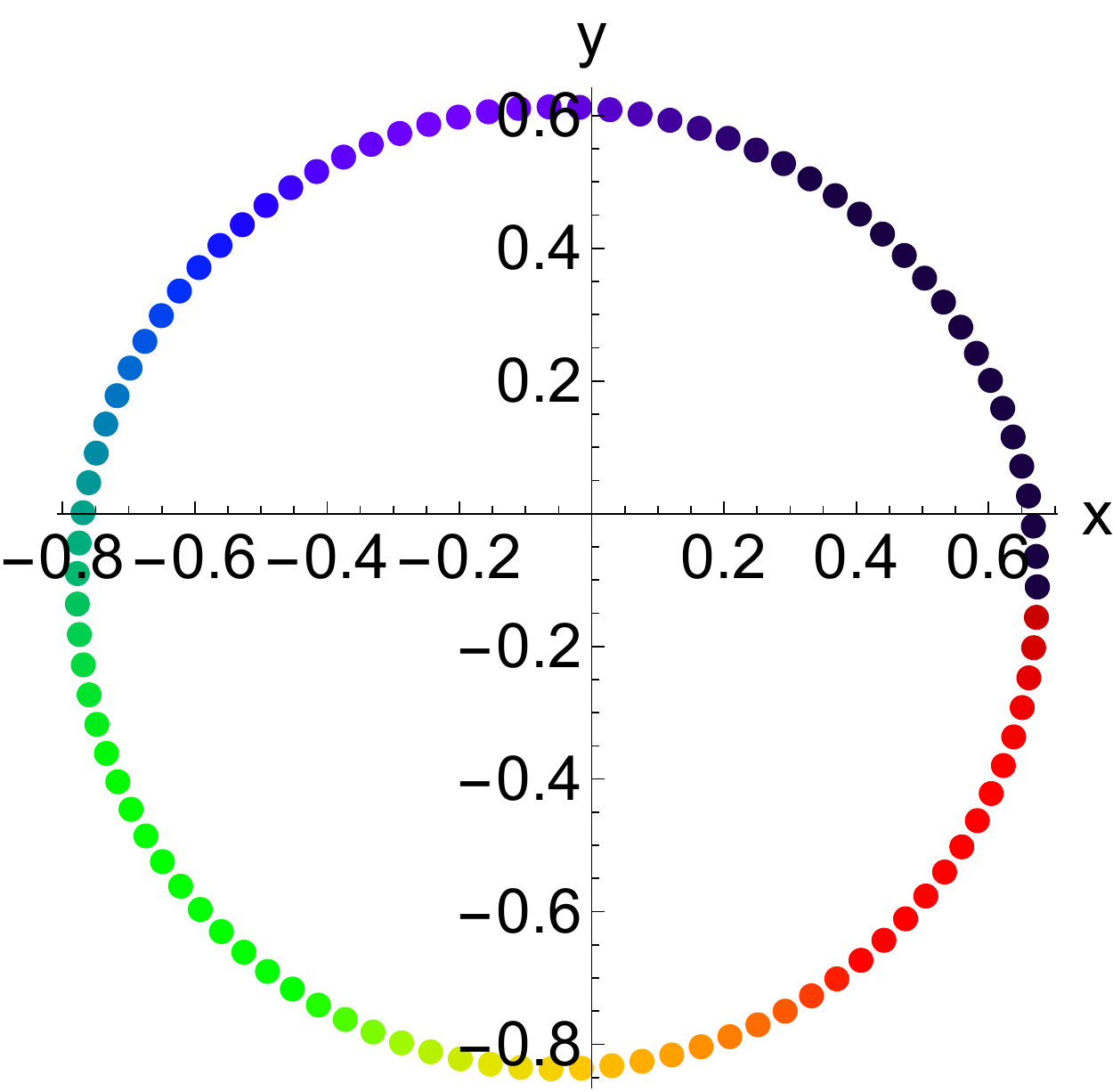}
 \caption{\textbf{Ring phase state.} Swarmalators are represented by colored dots in the $(x,y)$ plane where the color of each dot represents its phase. The state is found by numerically integrating the governing equations in \cite{o2018ring} with $(J_1, J_2, K, N) = (0,0.8,0,100)$.}
\label{ring}
\end{figure}

Some extensions of the work in \cite{o2017oscillators} have been carried out. One addresses the conspicuously empty upper right quadrant in the $(J,K)$ parameter plane in Figure~\ref{phase_diagram_model1}, where only the trivial static sync state appears. By adding phase noise to \eqref{theta_eom_model}, Hong discovered \cite{hong2018active} the active phase wave exists for $K > 0$. Unexpectedly, the splintered phase wave was not observed. O'Keeffe, Evers, and Kolokolnikov \cite{o2018ring} have extended \eqref{x_eom_model} by allowing phase similarity to affect the spatial repulsion term, as well as the spatial attraction term (i.e.~they multiply the second term in \eqref{x_eom_model} by a term $1+J_2 \cos(\theta_j - \theta_i)$). This led to the emergence of ring states, an example of which is depicted in Figure~\ref{ring}. They constructed and analyzed the stability of these ring states explicitly for a population of given size $N$. Analytic results for arbitrary $N$ are potentially useful for robotic swarms, which presumably are realized in the small $N$ limit. Another offshoot of this analysis was a heuristic method to predict the number of clusters which form in the splintered phase wave state; they viewed each cluster of synchronized swarmalators as one giant swarmalator which let them re-imagine the splintered phase wave state as a ring, allowing them to leverage their analysis. A precise description of the number of clusters formed is an open problem. 

\textbf{Iwasa-Tanaka model}. Iwasa and Tanaka proposed and studied a `swarm-oscillator' model in a series of papers \cite{tanaka2007general, iwasa2011juggling, iwasa2017mechanism, iwasa2010dimensionality, iwasa2012various}. The inspiration for their work comes from chemotactic oscillators, i.e., oscillators moving around in a diffusing chemical which mediates their interactions. They began with the general~model
\begin{align}
    \dot{\mathbf{X}}_i(t) &= f(\mathbf{X}_i) + k g(S(\mathbf{r}_i,t)) \\
    m \ddot{r_i}(t) &= - \gamma \dot{r}_i - \sigma(\mathbf{X}_i) \nabla S \\
    \tau \partial_{\tau} S(\mathbf{r}, t) &= -S + d \nabla^2 S + \sum_i h(\mathbf{X}) \delta(\mathbf{r} - \mathbf{r}_i),
\end{align}
\noindent
where $\mathbf{X}_i$ represents the internal state (which will later be identified as a phase), $\mathbf{r}$ represents the position of the $i$-th oscillator, and $S$ represents the concentration of the background chemical. By means of a center manifold calculation and a phase reduction technique they derived the simpler equations
\begin{align}
    \dot{\psi}_i(t) &=  \sum_{j \neq i} e^{- |\mathbf{R}_{ji}|} \sin( \Psi_{ji} - \alpha |\mathbf{R}_{ji}| - c_1 ) \label{Tanaka_x} \\
    \dot{\mathbf{r}}_i(t) &= c_3 \sum_{j \neq i} \hat{\mathbf{R}}_{ji} e^{- |R_{ji}|} \sin( \Psi_{ji} - \alpha |\mathbf{R}_{ji}| - c_2 ) \label{Tanaka_theta}
\end{align}
\noindent
where $\mathbf{R}_{ji} = \mathbf{R}_j - \mathbf{R}_i$, $\Psi_{ji} = \psi_j - \psi_i$, and $\psi_i$ is the $i$-th oscillator's phase. We call this Iwasa-Tanaka model. 
Notice the space-phase coupling in this model is somewhat peculiar; in contrast to the swarmalator model given by \eqref{x_eom_model} and \eqref{theta_eom_model} the relative position $\mathbf{R}_{ji}$ and relative phase $\Phi_{ji}$ appear \textit{inside} the sine terms in both the $\dot{r}_i$ and $\dot{\psi}_i$ equations. Another difference between the two models is that $\dot{\mathbf{r}}_i$ in \eqref{Tanaka_x} has no hardshell repulsion term, which means the oscillators can occupy the same position in~space. 

The Iwasa-Tanaka model has rich collective states. An exhaustive catalogue of these states with respect to the model's four parameters is an ongoing effort \cite{iwasa2012various}. Highlights include a family of clustered states \cite{tanaka2007general,iwasa2010hierarchical,iwasa2010dimensionality} in which swarmalators collect in synchronous groups. The spatial distributions of these groups depend on their phase, similar to the splintered phase wave (Figure~\ref{stationary_states_2d}). The authors speculate this phase clustering is reminiscent of the chemotactic cell sorting during biological development \cite{tanaka2007general}. The Iwasa-Tanaka model also produces ring states \cite{iwasa2010hierarchical}, as well as an interesting `juggling' state \cite{iwasa2011juggling} in which the population forms a ``rotating triangular structure whose corers appear to `catch' and `throw' individual elements'' -- in other words, the population juggles the elements around the corners of a triangle (see Figure 1 in \cite{iwasa2011juggling}). Aside from theoretical novelty, this juggling could conceivably be exploited in robotic swarms, potentially allowing some form of relay between the elements.


\subsection{Alignment and synchronization}
In our proposed taxonomy, systems that combine alignment and synchronization are characterized by an internal phase $\theta$ and an orientation $\beta$ without a dynamic spatial degree of freedom. In other words, the particles' position $\mathbf{x}$ might affect their $\theta$ and $\beta$ dynamics, but $\mathbf{x}$ itself does not evolve in time. Although units characterized by just a phase $\theta$ and orientation $\beta$ might seem odd, Leon and Liverpool studied a system with units which meet these criteria: a class of soft active fluids which constitute a `new type of nonequilibrium soft matter -- a space-time liquid crystal' \cite{leoni2014synchronization}. They developed a phenomenological theory of these soft fluids which allowed them to derive dynamical equations for order parameters quantifying the orientation order, phase order, and orientation-phase order. These revealed collective states which maximize each of these order parameters: aligned states with orientational order but no phase order, sync'd states with phase order but no orientational order, and states with both phase and orientational order. They were able to partially analyze these states, and conjectured the states could be realized in protein-filaments mixtures, such as cell cytoskeletons \cite{toner1995long,ahmadi2006hydrodynamics} or tissue-forming cells \cite{sachs2005pattern}. 


\subsection{Alignment, aggregation, and synchronization}
Systems with units that align, aggregate, and synchronize -- and therefore have dynamic state variables $\mathbf{x}, \theta, \beta$ -- are the least well studied. A small study was carried out in \cite{o2017oscillators}, with the aim of checking if the swarmalator states reported are generic, i.e., robust to the inclusion of alignment dynamics (it was found they were). Beyond this preliminary study, the space of possible behaviors arising from alignment, aggregation, and synchronization is largely unexplored.


\subsection{Alignment and aggregation}
While not strictly within our proposed taxonomy of swarmalator systems, we give a brief review of studies on aggregation and alignment. We do this because alignment can be viewed as a type of synchronization, where instead of units adjusting to a common phase $\theta$, units instead adjust a common orientation $\beta$.  Or put another way, because an internal phase $\theta$ and orientation $\beta$ are formally equivalent -- both being circular variables -- particles aligning their orientations is analogous to oscillators aligning their phases. 

Vicsek set the paradigm of this class of models with a beautiful and simple model:
\begin{align}
& \mathbf{r}_i(t + \delta t)= \mathbf{r}_i(t) + (\nu \delta t) \hat{n} \\ 
& \beta_i(t + \delta t) = \langle \beta_j \rangle_{|r_j - r_i|<r} + \eta_i(t) \label{vicsek},
\end{align}
\noindent
where $\hat{n} = (\cos(\beta), \sin(\beta))$, $\eta_i(t)$ is a white noise, $\nu$ is the particles' speed, and $r$ is the coupling range. For sufficiently strong noise this model exhibits a flocking transition, where particles switch from erratic incoherent movements to moving in a unidirectional flock.

Leibchen and Levis \cite{liebchen2017collective, levis2018activity} modified the Vicsek model as follows:
\begin{align}
&\dot{\mathbf{r}}_i= \nu \vec{n}_i \\ 
& \dot{\beta_i} = \omega_i + \frac{K}{\pi R^2} \sum_{j \in \partial_i}^N \sin(\theta_j - \theta_i) + \sqrt{2D} \eta_i,
\end{align}
\noindent
where $\partial_i$ is the set of neighbours of the $i$-th particle, $K$ is the phase coupling, and $D$ is the noise strength. The key new feature is the $\omega_i$ term, signifying the particles have an intrinsic rotation. They find three collective states: a disordered state in which neither spatial order nor phase order exists, a chiral state in which macroscopic `droplets' of synchronized oscillators form in a sea of otherwise desynchronized oscillators, and finally a \textit{mutual flocking} state in which oscillators with `opposite chirality cooperate to move coherently at a relative angle, forming non-rotating flocks'. The theoretical novelty of the latter state is that -- in contrast to regular, non-moving, locally-coupled oscillators coupled in two or three dimensions -- long-range synchronous clusters are observed, despite the fact that the oscillators are only locally coupled.


\section{Swarmalators in the real-world}

To our knowledge, there are just two works which unequivocally realize swarmalators in the real world -- by this we mean, precisely, a real-world system in which a bidirectional space-phase coupling is unambiguously exhibited. These works are: 

\textbf{Swarmalatorbots}. Bettstetter et al first realized the collective states of the swarmalator model (Figure~\ref{stationary_states_2d}) in the lab using small robots \cite{gniewek19}. They programmed `swarmalatorbots' whose governing equations were derived from the swarmalator model \eqref{x_eom_model}, \eqref{theta_eom_model}.

\textbf{Magnetic domain walls}. To our knowledge, the first realization of a natural swarmalator system was found by Hrabec et al when studying the magnetic domain walls \cite{hrabec2018velocity}. Ordinarily, domain walls are described by a single spatial degree of freedom $\mathbf{x}$. But as the authors note, the dynamics on the walls also depend on their internal structure. The authors minimally describe this internal state by a one dimensional polarization angle of the internal magnetic field $\theta$. This allows the wall to be viewed as a point particle with position $\mathbf{x}$ and phase $\theta$ -- to be viewed as a swarmalator \cite{hrabec2018velocity}. An experimental study of  coupling between two domain walls -- usually just one wall is studied -- revealed the walls can synchronize, which in turn affects the walls' velocity. Richer space-phase dynamics, such as families of Lissajous curves, are also reported. 

\begin{center}
***
\end{center}

Beyond swarmalatorbots and magnetic domain walls, there are many systems in which a bidirectional coupling between swarming and synchronization might exist. We list these candidate swarmalator systems below:

\textbf{Myxobacteria}. Myxobacteria are bacteria commonly found in soil, and can produce interesting collective effects \cite{igoshin2001pattern}. The phase variable of myxobacteria characterizes the internal growth cycle of the cell. Groups of cells interact through cell-to-cell contact, theorized to provide a channel through which swarming and synchronization can couple. Populations of myxobacteria can exhibit a `ripple phase', in which complex patterns of waves propagate through population \cite{igoshin2001pattern}. In contrast to other wave phenomena in biological systems, such as the well-studied Dictyostelium discoideum, these rippling waves do not annihilate on collision. In \cite{igoshin2001pattern} a Fokker-Planck type equation was used to analyze rippling waves. Realizing them in a microscopic swarmalator system is an open problem.

\textbf{Biological microswimmers}. `Microswimmers' is an umbrella term for self-propelled micro\-organisms confined to fluids, such as celia, bacteria, and sperm \cite{elgeti2015physics}. Groups of micro\-swimmers show swarming behavior as a result of cooperative goal seeking, such as searching for food or light. They can also synchronize: Here the phase variable is associated with the rhythmic beating of swimmers' tails which -- through hydrodynamic interactions -- can synchronize with the beatings of others swimmers' tails. Whether or not this hydrodynamics provides a bidirectional coupling between sync and swarming -- as required of swarmalators -- is unclear, but to us seems plausible. Researchers have developed models of sperm under this assumption \cite{yang2008cooperation} and found clusters of synchronized sperm consistent with real data~\cite{hayashi1996insemination}. Vortex arrays of sperm have also been reported \cite{riedel2005self}. Here, sperm self-organize into subgroups arranged in a lattice; within each subgroup, the sperm move in a vortex, wherein a correlation between the their angular velocity and their phase velocity is realized, reminiscent of the splintered phase wave state (Figure~\ref{stationary_states_2d}). Simulations of realistic models have been carried out \cite{belovs2017synchronized} which show other interesting~phenomena.  

\textbf{Japanese tree frogs}. During mating season, male Japanese tree frogs attract females by croaking rhythmically. The croaking of neighbouring frogs tend to anti-synchronize due to a precedence effect: croaking shortly after a rival makes a frog look less dominant. Researchers have theorized that this competition leads to mutual interaction between frogs' space and phase dynamics, coupling sync and swarming. Models based on these assumption produce ring-like states where frogs arrange themselves on the borders of fields with interesting space-phase patterns, some of which are consistent with data on real tree-frogs' behavior collected in the wild.

\textbf{Magnetic colloids}. Similar to magnetic domain walls, a colloid's constituent particles can synchronize the orientation of their magnetic dipole vector when sufficiently close. When in solution, they are free to move around in space, creating a feedback loop between their space and phase dynamics. In ferromagnetic colloids confined to liquid-liquid interfaces, Snezhko and Aranson \cite{snezhko2011magnetic} found this interaction leads to the formation of `asters' -- star-like arrangements of particles whose spatial angles correlate with the orientations of their magnetic dipole vectors, equivalent to the static phase wave (Figure~\ref{stationary_states_2d}). Yan et al explored how synchronization is useful in colloids of Janus particles \cite{yan2012linking}. Janus particles are micrometer-sized spheres with one hemisphere covered with nickle which gives them non-standard magnetic properties. In particular when subject to a precessing magnetic field, they oscillate about their centers of mass. This oscillation creates a coupling between particles, giving rise to `synchronization-selected' self-assembly. For example, zig-zag chains of particles and microtubes of synchronized particles can be realized.


\section{Swarmalators in bio-inspired computing}
The computer engineering community has been inspired by coupled oscillators to develop new techniques for synchronization in communication and sensor networks~\cite{mathar96,hong05,werner-allen05,simeone08,babaoglu2007firefly,tyrrell10}. Here, instead of  Kuramoto-type oscillators \eqref{kuramoto}, models of pulse-coupled oscillators have been borrowed. As the name suggests, pulse-coupled oscillators communicate by exchanging short signals; this time-discrete variant of coupling is a more natural fit for application in technology, where smooth, continuous coupling -- as exemplified by Kuramoto oscillators --  is costly to achieve.
The canonical model of pulse-coupled oscillators is the Peskin model \cite{peskin1975mathematical}, defined by \begin{equation}
    \dot{x_i} = S_0 - \gamma x_i
    \label{peskin}
\end{equation}
\noindent
with $S_0, \gamma > 0$ and $x_i$ is a voltage like state variable for the $i$-th oscillator. When the oscillators' voltage reach a threshold value they ($i$)~fire a pulse which instantaneously raises the voltage of all the other oscillators ($ii$)~reset their voltage to zero, along with any other oscillator whose voltage was raised above the threshold on account of receiving a pulse. The synchronization properties of the Peskin model are well-studied \cite{bottani1995pulse,mirollo1990synchronization}. More recently, estimates for the convergence speed have been derived~\cite{o2015synchronization,o2016transient}.
 
The simplicity, distributed nature, adaptability, and scalability of pulse-coupled oscillators make them attractive from an engineering point of view, where temporal coordination is often a goal. For example, synchronization is required for many tasks in different layers of computing and communications systems~\cite{bregni02} such as in the alignment of transmission slots for efficient medium access~\cite{Roberts:1975:APS}, scheduling of sleep cycles for energy efficiency~\cite{ye04}, and coordination of sensor readings to capture a scene from different perspectives~\cite{aghajan:book}. It is the relative synchrony of the networked units, not necessarily the absolute time, that is relevant here. The synchronization precision required is determined by the specific task. The achievable precision depends on the environment and hardware, and is influenced by deviations in delays and phase rates. Experiments in laboratory environments show that the precision is in the order of hundred microseconds with low-cost programmable sensor platforms \cite{werner-allen05,pagliari11} and a few microseconds with field-programmable gate array (FPGA)-based radio boards~\cite{brandner16:cn}. Further use cases can be found in acoustics (synchronizing multiple loudspeakers) and energy systems (synchronizing decentralized grids \cite{rohden12,skardal15}), to give two examples. 

Synchronization is also important in robotics in order to perform coordinated movements -- and this is where temporal and spatial coordination unite. As mentioned, Bettstetter and colleagues~\cite{gniewek19} extended the swarmalator model \cite{o2017oscillators} so that it could be applied to mobile robots. They implemented the extended model in the Robot Operating System~2 (ROS\,2) and experimentally demonstrated that the space-time patterns achieved in theory (Figure\,\ref{stationary_states_2d}) can be reproduced in the real world. Beyond realizing these specific states, we conjecture that swarmalator-type models will enable novel self-assembly procedures in other robotics groups, in turn enabling collaborative actions in monitoring, exploration, and manipulation. As envisioned in \cite{gniewek19}, underwater robots could be designed, which -- in imitation of biological microswimmers -- could both swim in formation and synchronize their fin movements, thereby enhancing their functional capabilities. 


More speculatively, the swarmalator concept could be useful in the design of autonomous transport systems; for example, when multiple vehicles driving in a convoy have to avoid collisions with other convoys due to crossings and for the purpose of overtaking. The model could also enable self-configuring distributed antenna arrays (or loudspeakers) in which multiple antenna elements (or sound sources) automatically arrange their positions and orientations to create specific radiation patterns used to send radio (audio) signals in a synchronous way. Another promising application field is the planning and replanning of processes in factories, where products and machines must follow a certain space-time order and where standard optimization methods reach their limits due to the high system complexity. A final, more playful, application is in art: Artistic aerial light shows created by drone swarms running the swarmalator model produce charming visual displays. 



\section{Discussion}
The implicit strategy in the theoretical works we  reviewed is: given a model, which spatiotemporal patterns emerge? Future work could consider the inverse strategy: given a desired spatiotemporal pattern, which form should the model take on? -- a question often considered in engineering contexts. From a more general perspective, one has to design the local rules and interactions that guide a self-organizing system toward a desired global state \cite{prehofer05:commag}. Von Brecht et al have studied this inverse question for swarming systems~\cite{von2012soccer}; perhaps their tools could be extended to swarmalators. 

Future work could also extend the reviewed models, which -- recall -- were designed to be as minimal as possible. A wealth of synchronization phenomena have been found by adorning the Kuramoto model with new features like mixed-sign coupling \cite{hong2011kuramoto,maistrenko2014solitary,hong2016phase, hong2016correlated}, non-local coupling \cite{abrams2004chimera, abrams2008solvable}, and delayed interactions \cite{yeung1999time}. Equipping swarmalators with these features would likely cure the poverty of phenomena in the first, third, and fourth quadrants of the $(J,K)$ plane (Figure~\ref{phase_diagram_model1})  and produce new dynamics in Iswasa-tanaka and Vicesek type models, too. Phase models more sophisticated than the Kuramoto model could also be explored; the Winfree model \cite{winfree67} or the newly introduced Janus oscillators \cite{nicolaou2019multifaceted} would be exciting to experiment with. Swarmalators with discrete and spatially local coupling also merit study and are useful in computer and communication systems. Stitching the Peskin model \eqref{peskin} to the aggregation equation \eqref{agg} or the Vicsek model \eqref{vicsek} seems the natural way to do this. Continuing to study finite systems of swarmalators \cite{o2018ring} is also pertinent, since many robotic systems lie in the low $N$ regime. And finally, beyond richer phase models, the spatial dynamics of swarmalators could also be generalized. Even within the framework of the aggregation equation~\eqref{agg}, a menagerie of spatial patterns have been catalogued \cite{kolokolnikov2011stability}. Swarmalator counterparts to these states would be exciting to explore. 

We hope to have outlined that swarmalators have great potential in computing and other fields of technology. Here we see swarmalators only as a case study in a broader class of systems with large technological utility: systems whose units have both spatial and internal degrees of freedom. The one-dimensional phase $\theta$ of a swarmalator is perhaps the simplest instance of an internal degree of freedom, which can more generally be represented by a feature vector~$\mathbf{f}$. This vector could describe a particle's (three dimensional) magnetic or electric field; a person's political affiliation, mood, or health state; a bacteria's phase in a non-circular developmental cycle; and the mode of a robot, machine, or vehicle. The collective states arising when particles' features $\mathbf{f}$ and positions $\mathbf{x}$ defines a wide, scarcely investigated area. We anticipate an exploration of this area will bear rich fruit in both Nature and technology. \\


\section*{Acknowledgements}
The work of C.\,Bettstetter is supported by the Austrian Science Fund (FWF) [grant P30012] and the Popper Kolleg {\it Networked Autonomous Aerial Vehicles}.

\bibliographystyle{apsrev}

\end{document}